\documentclass[english,reqno]{article}
\usepackage{amsmath}
\usepackage{cite}
\usepackage{graphicx}
\usepackage{amsfonts}
\usepackage{amssymb}
\usepackage{times,macros}
\newcommand{\beq}{\begin{equation}}
\newcommand{\eeq}{\end{equation}}

\newcommand{\ra}{\rightarrow}

\newcommand{\ti}{\times}

\newcommand{\fr}[2]{{\textstyle \frac{#1}{#2} }}

\newcommand{\bra}{\langle}

\newcommand{\ket}{\rangle}
\newcommand{\al}{\alpha}

\newcommand{\be}{\beta}

\newcommand{\Ga}{\Gamma}
\newcommand{\de}{\delta}
\newcommand{\De}{\Delta}

\newcommand{\si}{\sigma}
\newcommand{\up}{\Upsilon}

\newcommand{\bz}{\bar{z}}

\newcommand{\CF}{{\mathcal F}}

\newcommand{\CH}{{\mathcal H}}

\newcommand{\CV}{{\mathcal V}}
\newcommand{\CW}{{\mathcal W}}

\newcommand{\BR}{{\mathbb R}}

\newcommand{\BS}{{\mathbb S}}

\newcommand{\BZ}{{\mathbb Z}}

\def\my6j#1#2#3#4#5#6{
\big\{\,{}^{#1}_{#2}\;{}^{#3}_{#4}\,|\,{}^{#5}_{#6}\,
\big\}_{b}}

\def\new6j#1#2#3#4#5#6{
\big\{\,{}^{#1}_{#2}\;{}^{#3}_{#4}\,|\,{}^{#5}_{#6}\,
\big\}_{b}'}

\def\fus#1#2#3#4#5#6{
F_{#5#6}\!\big[\, {}_{#4}^{#3}\;{}_{#1}^{#2}\,
\big] }

\def\Sp{
\mathbb{S}
}

\def\al{
\alpha
}

\def\Ga{
\Gamma_{b}
}

\def\up{\Upsilon_{b}}

\def\uq{\mathcal{U}_q(\mathfrak{sl}(2,\mathbb{R}))}

\begin{document}
%\bigskip
%\hfill\begin{minipage}{3cm}
%SfB 288 Preprint\\ AEI 2001-124
%\\
%LPM 01/32
%\end{minipage}
%\vspace{0.8cm}
% \title[Boundary Three Point Function]{Boundary Liouville Field Theory:\\
\title{Boundary Liouville Field Theory:\\Boundary Three Point Function}

\author{B.Ponsot and J.Teschner}

\begin{abstract}
Liouville field theory is considered on domains
with conformally invariant boundary
conditions. We present an explicit expression for
the three point function of boundary fields in 
terms of the fusion coefficients which determine the monodromy
properties of the conformal blocks.
\end{abstract}
% \thanks{${}^{\dagger}$ Max-Planck-Institut f\"ur Gravitationsphysik, 
% Albert Einstein Institut, Am M\"uhlenberg 1, 14476 Golm, Germany,\\
% ${}^{\flat}$Institut f\"ur 
% theoretische Physik, Freie Universit\"at Berlin, Arnimallee 14, 14195 
% Berlin, Germany}\maketitle
\address{Max-Planck-Institut f\"ur Gravitationsphysik, 
Albert Einstein Institut, Am M\"uhlenberg 1, 14476 Golm, Germany,\\
Institut f\"ur 
 theoretische Physik, Freie Universit\"at Berlin, Arnimallee 14, 14195 
Berlin, Germany}
\maketitle

\section{Introduction}

Liouville theory seems to be a universal building block
that appears in various contexts such as noncritical 
string theory, two-dimensional gravity or D-brane physics. 
It is also closely related to the $SL(2)$ or $SL(2)/U(1)$ WZNW
models which are interesting as solvable models for 
string theory on noncompact curved backgrounds. From a more 
general point of view one may regard Liouville theory
as a prototype for an interacting conformal field with noncompact
target space. It should therefore serve as a natural starting
point for the development of techniques for the exact solution
of such conformal field theories.

In the case of Liouville theory with periodic boundary conditions
we now have a relatively complete characterization \cite{Te}: Knowledge
of the spectrum of the theory and three point functions of primary
fields allow one to consistently reconstruct arbitrary expectation values
of local fields on the sphere or cylinder. 

Our understanding is less satisfactory in the case of Liouville
theory on two-dimensional domains with boundary such as the infinite strip,
the upper half plane or the disk: One would again expect the theory to be 
fully characterized in terms of a finite set of structure functions
together with the knowledge of the spectrum of the theory on the 
strip. Consistency of the reconstruction of the theory from these 
fundamental data requires them to satisfy consistency conditions 
very similar to those formulated by Cardy and Lewellen \cite{CL}
in the case of rational conformal field theories. A part of these
data has been determined and some of the basic consistency
conditions have been verified \cite{FZZ}\cite{tesch1}\cite{hosomichi}.
What is missing are the determination of the three point function 
of boundary operators and the verification that these data satisfy
the full set of conditions ensuring consistency of the reconstruction 
of the theory. The aim of the present paper is to propose an 
explicit expression for the three point function 
of boundary operators as the solution to one of the most important 
consistency conditions expressing the associativity of the 
product of boundary operators.

The structure of this paper is as follows: 
The following section gathers those results on Liouville theory
that we will use in the present paper. 
The third section then contains our proposal for the three point function 
of boundary operators. It is based on the observation \cite{runkel}
that an ansatz for that three point function in terms of the fusion
coefficients naturally leads to a solution of the 
consistency condition that expresses the associativity of the 
product of boundary operators. It remains to fix the remaining freedom 
by imposing certain normalization conditions.

Some concluding remarks are made in section 4, and the appendices 
contain some technical points used in the main text.

\section{Requisites}

\subsection{Liouville theory on the sphere}

Let us begin by recalling some results on Liouville theory that 
will be relevant for the subsequent discussion, see \cite{Te} for more
details and references:

LFT on the sphere is semiclassically defined by the
following action
\begin{eqnarray}
\mathcal{A}_{L}=\int\left(\frac{1}{4\pi}(\partial_a\phi)^{2}+\mu e^{2b\phi}\right)d^{2}x, 
\label{action Liou}
\end{eqnarray}
with the following boundary condition on the Liouville field $\phi$
\begin{equation}
\phi(z,\bar{z})=-Q\log(z\bar{z})+O(1)\ \ \ \ \ \mathrm{at}\ \ |z|\rightarrow
\infty.\label{sphere}%
\end{equation}
The parameter $b$ is related to Planck's constant $\hbar$ via 
$b^2=\hbar$, the scale parameter
$\mu$ is often called the cosmological constant, and $Q$ is the background
charge
$$ Q=b+1/b.$$

It was first proposed in \cite{curtright}
that Liouville theory can be quantized as a conformal field theory 
with a space of states that decomposes as follows into irreducible
unitary highest weight representations $\CV_{\al}$ of the Virasoro algebra:
\begin{equation}\label{spec}
\mathcal{H}=\int_{\Sp} d\al\; \CV_{\al}\otimes \CV_{\al},\qquad
\Sp=\frac{Q}{2}+i\mathbb{R}^{+}.
\end{equation}
The highest weight $\De_{\al}$ of the representation $V_{\al}$ 
was parametrized as $
\Delta_{\alpha}=\alpha(Q-\alpha)$. The action of the Virasoro 
algebra on $\CH$ is generated by the modes of the energy momentum tensor:
\begin{eqnarray}
T(z)&=&-(\partial\phi)^{2}+Q\partial^{2}\phi, \nonumber \\
\bar{T}(\bar{z})&=&-(\bar{\partial}\phi)^{2}+Q\bar{\partial}^{2}\phi. \nonumber
\end{eqnarray}
The central charge of the Virasoro algebra is then given in terms of $b$ 
via
$$ c_{L} = 1 + 6Q^{2}. $$

The local observables
can be generated from the 
fields $V_{\al}(z,\bar{z})$
which semiclassically ($b\rightarrow 0$) correspond to 
exponential functions 
$e^{2\al\phi(z,\bar{z})}$ of the Liouville field. 
The fields $V_{\al}(z,\bar{z})$ transform as primary fields under 
conformal transformations with conformal weight $\De_{\al}$.
Thanks to conformal symmetry,
the fields $V_{\al}(z,\bar{z})$ are fully characterized by the 
three point functions 
$$
C(\al_3,\al_2,\al_1)=\lim_{z_3\ra\infty}|z_3|^{4\De_{\al_3}}
\langle 0|V_{\al_3}(z_3,\bar{z}_3)V_{\al_2}(1,1)
V_{\al_1}(0,0)|0\ket.$$
An explicit formula for the three point function was proposed in
 \cite{DO,AAl}\footnote{see the 
Appendix A for some definitions and properties of the
special functions used in this article}
\begin{equation}\label{3pointsZZ}\begin{aligned}
C(\al_3, & \al_2,\al_1)= \left\lbrack \pi \mu \gamma(b^{2})b^{2-2b^{2}}
\right\rbrack^{\frac{1}{b}(Q-\sum_{i=1}^3\al_i)}  \\
&  \frac{\Upsilon_0\up(2\al_1)\up(2\al_2)\up(2\al_3)}{\up(\al_1+\al_2+\al_3-Q)\up(\al_1+\al_2-\al_3)\up(\al_1+\al_3-\al_2)\up(\al_2+\al_3-\al_1)}, 
\end{aligned}\end{equation}
where $\gamma(x)=\frac{\Gamma(x)}{\Gamma(1-x)}, \Upsilon_{0}=\text{res}_{x=0}\frac{d\up(x)}{dx}$.\\
These pieces of information indeed amount to a full characterization
of Liouville theory on the sphere or cylinder:
Multipoint correlation functions can be factorized into 
three point functions by summing over intermediate states.
Let us consider as prototypical example the four point function
$\langle 0|\prod_{i=1}^4 V_{\al_i}(z_i,\bar{z}_i)|0\ket$. 
Such four point functions may be represented by summing over intermediate 
states from the spectrum (\ref{spec}) iff the variables $\al_4,\dots,\al_1$
are restricted to the range\footnote{It turns out \cite{Te} that the four-point function defined 
in the range (\ref{funran}) 
permits a meromorphic continuation to generic values
of $\al_4,\dots,\al_1$.}
\begin{equation}\label{funran}\begin{aligned}
 2|\text{Re}(\al_1+\al_2-Q)|<Q,     &\qquad 2|\text{Re}(\al_1-\al_2)|<Q,  \\
 2|\text{Re}(\al_3+\al_4-Q)|<Q,     &\qquad 2|\text{Re}(\al_3-\al_4)|<Q. 
\end{aligned}\end{equation}
Inserting a complete set of intermediate states between $\bra 0|V_{\al_4}
V_{\al_3}$ and $V_{\al_2}V_{\al_1}|0\ket$ would lead to an expression 
of the following form:
\begin{equation}\begin{aligned}\label{s-channel}
\bra 0| V_{\al_4} & (z_4,\bar{z}_4)  V_{\al_3}(z_3,\bar{z}_3)V_{\al_{2}}(z_2,\bar{z}_2)V_{\al_1}(z_1,\bar{z}_1)|0\ket =  \\
&=\;\int_{0}^{\infty}dP\;
C(\al_4,\al_3,Q/2-iP)C(Q/2+iP,\al_{2},\al_{1})
|\mathcal{F}^{s}(\Delta_{\al_i},\Delta,z_i)|^{2} 
\end{aligned}\end{equation}
$\mathcal{F}^{s}(\Delta_{\al_i},\Delta,z_i)$ is the s-channel
conformal block which is completly determined by the conformal
symmetry (although no closed formula is known for it in general).
\begin{align}\label{block}
\mathcal{F}^{s}(\Delta_{\alpha_{i}},\Delta,z_{i}) = \; &
  (z_{4}-z_{2})^{-2\Delta_{2}}(z_{4}-z_{1})^{\Delta_{2}+\Delta_{3}%
-\Delta_{1}-\Delta_{4}}(z_{4}-z_{3})^{\Delta_{1}+\Delta_{2}-\Delta_{3}%
-\Delta_{4}}\nonumber\\
& \ti (z_{3}-z_{1})^{\Delta_{4}-\Delta_{1}-\Delta_{2}-\Delta_{3}}
   \mathcal{F}_P\!\left[{}_{\al_4}^{\al_3}\;{}_{\al_1}^{\al_2}{}\right]\!
(\eta) \nonumber
\end{align}
where
$
\eta=\frac{(z_{1}-z_{2})(z_{3}-z_{4})}{(z_{2}-z_{4})(z_{1}-z_{3})}
$
and $\Delta_{\al_{i}}=\al(Q-\al)$, $\Delta=\frac{Q^2}{4}+ P^2$.
Locality of the fields $V_{\al}$ or associativity of the 
operator product expansion would lead to an alternative
representation for $\langle 0|\prod_{i=1}^4 V_{\al_i}(z_i,\bar{z}_i)|0\ket$
as sum over {\it t-channel} conformal blocks $\CF^t$: 
\begin{equation}\label{t-channel}\begin{aligned}  
\bra 0| V_{\al_4} & (z_4,\bar{z}_4)  V_{\al_3}(z_3,\bar{z}_3)V_{\al_{2}}(z_2,\bar{z}_2)V_{\al_1}(z_1,\bar{z}_1)|0\ket= \\
&=  \int_{0}^{\infty}dP \;
C(\al_4,Q/2-iP,\al_1)C(Q/2+iP,\al_{3},\al_{2})
|\mathcal{F}^{t}(\Delta_{\al_i},\Delta,z_i)|^{2}
\end{aligned}\end{equation}
For the 
equivalence of the two representations (\ref{s-channel}) and
(\ref{t-channel}) it is crucial that 
there exist \cite{Te} invertible fusion transformations between s-
and t-channel conformal blocks, defining the fusion coefficients:
\begin{eqnarray}
\mathcal{F}^{s}(\Delta_{\al_i},\Delta_{\al_{21}},z_i)
\;=\;\int_{\Sp}d\al_{32}\;
F_{\al_{21}\al_{32}}
\!\left[\,{}_{\al_4}^{\al_3}\;{}_{\al_1}^{\al_2}{}\,\right]
\,\mathcal{F}^{t}(\Delta_{\al_i},\Delta_{\al_{32}},z_i).
\label{transfost}
\end{eqnarray}
In \cite{PT1}, an explicit formula for the fusion
coefficients was proposed in
terms of the Racah-Wigner
coefficients for an appropriate continuous series of 
representations of the quantum group $\uq$ with deformation parameter 
$q=e^{i\pi b^{2}}$. This formula was subsequently \cite{Te} 
confirmed by direct calculation.
The resulting expression for the fusion coefficients is the following:
\[\begin{aligned}
{} & F_{\sigma_{2}\beta_{3}}
\!\left[\,{}_{\si_3}^{\be_2}\;{}_{\si_1}^{\be_1}{}\,\right]= \\
&
\frac{\Ga(2Q-\beta_1-\beta_2-\beta_{3})\Ga(\beta_2+\beta_{3}-\beta_1)\Ga(Q+\beta_2-\beta_1-\beta_{3})\Ga(Q+\beta_3-\beta_2-\beta_1)}{\Ga(2Q-\sigma_1-\beta_1-\sigma_{2})\Ga(\sigma_1+\sigma_{2}-\beta_1)\Ga(Q-\beta_1-\sigma_{2}+\sigma_1)\Ga(Q-\beta_1-\sigma_1+\sigma_{2})}  \\
&\ti
\frac{\Ga(Q-\beta_{3}-\sigma_1+\sigma_3)\Ga(\beta_{3}+\sigma_{1}+\sigma_{3}-Q)\Ga(\sigma_1+\sigma_3-\beta_{3})\Ga(\sigma_3+\beta_{3}-\sigma_1)}{\Ga(Q-\beta_{2}-\sigma_2+\sigma_3)\Ga(\beta_2+\sigma_2+\sigma_3-Q)\Ga(\sigma_2+\sigma_3-\beta_{2})\Ga(\sigma_{3}+\beta_2-\sigma_2)}  \\
& \ti\frac{\Ga(2Q-2\sigma_{2})\Ga(2\sigma_{2})}{\Ga(Q-2\beta_{3})\Ga(2\beta_{3}-Q)}
\frac{1}{i}\int\limits_{-i\infty}^{i\infty}ds \;\;
\frac{S_b(U_1+s)S_b(U_2+s)S_b(U_3+s)S_b(U_4+s)}
{S_b(V_1+s)S_b(V_2+s)S_b(V_3+s)S_b(Q+s)},  
\end{aligned}\]
where:
$$
\begin{array}{ll}
 U_1 = \sigma_{2}+\sigma_1-\beta_1 ,               
&  V_1 = Q+\sigma_{2}-\beta_{3}-\beta_{1}+\sigma_{3}, \\
 U_2 = Q+\sigma_{2}-\beta_1-\sigma_1,              
&  V_2 = \sigma_{2}+\beta_{3}+\sigma_{3}-\beta_1, \\
 U_3 = \sigma_{2}+\beta_{2}+\sigma_{3}-Q,          
&  V_3 = 2\sigma_{2}, \\
 U_4 = \sigma_{2}-\beta_{2}+\sigma_3. & 
\end{array}
$$
An important identity satisfied by the fusion coefficients is the 
so-called pentagon equation, which follows from a similar identity
satisfied by the Racah-Wigner coefficients mentioned previously 
\cite{PT2}.
\begin{equation}\begin{aligned}
\int_{\Sp}d\delta_{1}\; 
\fus{\al_1}{\al_2}{\al_3}{\beta_{2}}{\beta_{1}}{\delta_{1}}
\fus{\al_1}{\delta_1}{\al_4}{\al_5}{\beta_{2}}{\gamma_{2}}&
\fus{\al_2}{\al_3}{\al_4}{\gamma_{2}}{\delta_{1}}{\gamma_{1}} \\
& =\fus{\beta_1}{\al_3}{\al_4}{\al_5}{\beta_{2}}{\gamma_{1}}
\fus{\al_1}{\al_2}{\gamma_1}{\al_{5}}{\beta_{1}}{\gamma_{2}}.
\label{pentagone}
\end{aligned}
\end{equation}

\subsection{Liouville theory on domains with boundary}

One is also interested in understanding Liouville theory 
on a simply connected domain $\Gamma$ with a nontrivial boundary 
$\partial\Gamma$. 
For definiteness, we will only consider the conformally equivalent
cases where $\Gamma$ is either
the unit disk, the upper
half plane or the infinite strip.
  
Semiclassically, one may define the theory by means of the action
\begin{equation}
A_{\mathrm{bound}}=\int\limits_{\Gamma}\left(  \frac{1}{4\pi}(\partial_{a}%
\phi)^{2}+\mu e^{2b\phi}\right)d^{2}x +
\int\limits_{\partial\Gamma}\left(\frac{Qk}{2\pi}+
\mu_{B}e^{b\phi} \right) dx,\label{bound}%
\end{equation}
where $k$ is the curvature of the boundary $\partial\Gamma$ and 
$\mu_{B}$ is the so-called 
boundary cosmological constant. 
For the description of exact results 
in the quantum theory it was found to be useful \cite{FZZ} to 
parametrize $\mu_B$ by means of a variable $\si$ that is related to
$\mu_B$ via 
\begin{eqnarray}
\text{cos}2\pi
b\big(\sigma-\fr{Q}{2}\big)=\frac{\mu_{B}}{\sqrt{\mu}}\sqrt{\text{sin}(\pi
b^{2})}. \label{relation mu-sigma}
\end{eqnarray}
Requiring $\mu_{B}$ to be real one finds the two following regimes
for the parameter $\sigma$:
\begin{enumerate}
\item
if $\frac{\mu_{B}}{\sqrt{\mu}}\sqrt{\text{sin}(\pi b^{2})}>1$,
then $\sigma$ is of the form $\sigma= Q/2 + iP$
\item
if $\frac{\mu_{B}}{\sqrt{\mu}}\sqrt{\text{sin}(\pi b^{2})}<1$,
then $\sigma$ is real.
\end{enumerate}
Anticipating that all relevant objects will be found to possess meromorphic
continuations w.r.t. the boundary parameters $\si$, we shall discuss only
the first regime explicitly in the following. 

The Hamiltonian interpretation of the theory \cite{tesch1} is simplest
in the case that $\Gamma$ is the infinite strip. The associated
Hilbert space $\CH_B$ was found in \cite{tesch1} to decompose as follows
into irreducible representations of the Virasoro algebra:
\begin{equation}
\CH^{\rm B}\;=\;\int_{\BS}^{\oplus}d\be \;\CV_{\be}.
\end{equation}
The highest weight states generating the subrepresentation 
$\CV_{\be}$ in $\CH^{\rm B}$ will be denoted 
$|\be;\sigma_2,\sigma_1\ket$, where $\sigma_2$ ($\sigma_1$) 
are the parameters of the boundary conditions associated to the 
left (right) boundaries of the strip. It was proposed in 
\cite{FZZ}\cite{tesch1} that the states 
$|\be;\sigma_2,\sigma_1\ket$ satisfy a reflection relation
of the form 
\begin{equation}\label{reflprop}
|\be;\sigma_2,\sigma_1\ket=S(\be;\sigma_2,\sigma_1)
|Q-\be;\sigma_2,\sigma_1\ket.
\end{equation} which expresses the totally reflecting 
nature of the Liouville potential in (\ref{bound}). The 
following formula was given in \cite{FZZ} for the reflection 
coefficient $S(\be;\sigma_2,\sigma_1)$:
\begin{equation}\label{reflcoeff}\begin{aligned}
{} S(\beta_{3},\sigma_{3},\sigma_{1})
   = & (\pi\mu\gamma(b^{2})b^{2-2b^{2}})^{\frac{1}{2b}(Q-2\beta)}\ti\\
& \ti
\frac{\Gamma_{b}(2\beta_{3}-Q)}{\Gamma_{b}(Q-2\beta_{3})}
\frac{S_b(\sigma_{3}+\sigma_{1}-\beta_{3})S_b(2Q-\beta_{3}-\sigma_{1}-\sigma_{3})}{S_b(\beta_{3}+\sigma_{3}-\sigma_{1})
S_{b}(\beta_{3}+\sigma_{1}-\sigma_{3})}. 
\end{aligned}\end{equation}

In addition to the fields $V_{\al}(z,\bz)$ localized
in the interior of $\Gamma$, one may 
now also consider operators $\Psi_{\beta}^{\sigma_{2}\sigma_{1}}(x)$ 
that are localized at the boundary $\partial\Gamma$.
The insertion point $x$ may separate segments of the boundary 
with different boundary conditions $\sigma_{2}$ and $\sigma_{1}$.
The boundary fields $\Psi_{\beta}^{\sigma_{2}\sigma_{1}}(x)$ are
required to be primary fields with conformal
weight $\Delta_{\beta}=\beta(Q-\beta)$. They are therefore expected to
create states $|\be;\sigma_2,\sigma_1\ket$ and $\bra\be;\sigma_2,\sigma_1|$
via
\begin{equation}
\lim_{x\ra 0}
\Psi_{\beta}^{\sigma_{2}\sigma_{1}}(x)|0\ket
=|\be;\sigma_2,\sigma_1\ket,\quad
\lim_{x\ra \infty}
\bra 0|\Psi_{\beta}^{\sigma_{1}\sigma_{2}}(x)|x|^{2\De_{\be}}
=\bra Q-\be;\sigma_2,\sigma_1|.
\end{equation}

To fully characterize LFT on the upper half
plane, one needs to determine some 
additional structure functions beside the bulk three point
  function $C(\al_3,\al_2,\al_1)$:
\begin{enumerate}
\item Bulk one point function \cite{FZZ}:
\begin{equation}
\bra 0| V_{\alpha}(z,\bar{z})|0 \ket =\frac{U(\alpha|\si)}{\left|
z-\bar{z}\right|^{2\Delta_{\alpha}}}. \label{onepoint}%
\end{equation}
\item Boundary two point function \cite{FZZ}:
\begin{equation}\label{twopt}
\bra 0|
\Psi_{\beta_1}^{\sigma_{1}\sigma_{2}}(x)
\Psi_{\beta_2}^{\sigma_{2}\sigma_{1}}(0)
|0\ket
=\frac{\delta(\beta_2+\beta_1-Q)+
S(\beta_1,\sigma_{2},\sigma_{1})\delta(\beta_2-\beta_1)}{\left|
x\right|^{2\Delta_{\beta_1}}}.
\end{equation}
Let us remark that requiring the prefactor of the first delta-distribution
on the right hand side of (\ref{twopt}) to be unity
partially fixes the normalization of boundary operators.
The appearance of the second term in (\ref{twopt}) is a consequence of the 
reflection property (\ref{reflprop}).

\item bulk-boundary two point function\cite{hosomichi}:
\footnote{the bulk one point function is a special case of the
bulk-boundary coefficient with $\beta=0$.}
\begin{equation}
\bra 0| V_{\alpha}(z,\bar{z})\Psi_{\beta}^{\sigma\sigma}(x)|0\ket 
=\frac{R(\alpha,\beta
|\sigma)}{\left|  z-\bar{z}\right|  ^{2\Delta_{\alpha}-\Delta_{\beta}}\left|
z-x\right|  ^{2\Delta_{\beta}}} \label{bbound}%
\end{equation}
\item Boundary three point function:
\begin{equation}\begin{aligned}
\bra 0|
\Psi_{\beta_{3}}^{\sigma_{1}\sigma_{3}}(x_{3})
\Psi_{\beta_{2}}^{\sigma_{3}\sigma_{2}} & (x_{2})
\Psi_{\beta_{1}}^{\sigma_{2}\sigma_{1}}(x_{1})
|0\ket
=\\
&=
\frac{C_{\beta_{3}\beta_{2}\beta_{1}}^{\sigma_{3}\sigma_{2}\sigma_{1}}}{\left|
x_{21}\right|  ^{\Delta_{1}+\Delta_{2}-\Delta_{3}}\left|
x_{32}\right|
^{\Delta_{2}+\Delta_{3}-\Delta_{1}}\left|  x_{31}\right|  ^{\Delta_{3}%
+\Delta_{1}-\Delta_{2}}}. 
\end{aligned}\end{equation}
\end{enumerate}
Taking advantage of the 
reflection property (\ref{reflprop}), we shall consider instead of 
$C_{\beta_{3}\beta_{2}\beta_{1}}^{\sigma_{3}\sigma_{2}\sigma_{1}}$ the 
related quantity
\begin{equation}
C_{\beta_3|\beta_{2}\beta_{1}}^{\sigma_{3}\sigma_{2}\sigma_{1}}\;\equiv\;
C_{Q-\beta_{3},\beta_{2},\beta_{1}}^{\sigma_{3}\sigma_{2}\sigma_{1}}
\;\equiv\;
S^{-1}(\be_3;\si_1,\si_3)
C_{\beta_{3}\beta_{2}\beta_{1}}^{\sigma_{3}\sigma_{2}\sigma_{1}}.
\end{equation}
The present note will be devoted to the determination of this
last structure function.

\section{Boundary three point function}

\subsection{Associativity condition}

The basic consistency condition that the three-point function of boundary
operators has to satisfy expresses the associativity of the 
product of boundary fields.
Let us consider the 4 point function of boundary operators.
Inserting a complete set of intermediate states 
between the first two and the last two fields leads to an expansion
into conformal blocks of the following form:
\footnote{As in the discussion of the four point function
of bulk fields, we shall restrict
ourselves to the case where $Re(\beta_i)$, $i=1 \dots
4 $ are close enough to Q/2. In this case, $\beta_{21}$ is of the 
form $Q/2+iP$. It turns out a posteriori that the general case
can be treated by meromorphic continuation.}
\begin{equation}\label{4pointsBs}\begin{aligned}
\Bigl\langle 
\Psi_{Q-\beta_4}^{\sigma_{1}\sigma_{4}}(x_4)&
\Psi_{\beta_3}^{\sigma_{4}\sigma_{3}}(x_3)
\Psi_{\beta_2}^{\sigma_{3}\sigma_{2}}(x_2)
\Psi_{\beta_1}^{\sigma_{2}\sigma_{1}}(x_1)
\Bigr\rangle 
= \nonumber \\
&= 
\int_{\mathbb{S}}d\beta_{21}\;
C_{\beta_{4}|\beta_{3}\beta_{21}}^{\sigma_{4}\sigma_{3}\sigma_{1}}
C_{\beta_{21}|\beta_{2}\beta_{1}}^{\sigma_{3}\sigma_{2}\sigma_{1}}
\mathcal{F}^{s}(\Delta_{\beta_i},\Delta_{\beta_{21}},x_i). 
\end{aligned}\end{equation}
By using either cyclicity of correlation functions or 
associativity of the operator product expansion one would get a
second expansion (t-channel):
\begin{equation}\label{4pointsBt}\begin{aligned}
\Bigl\langle 
\Psi_{Q-\beta_4}^{\sigma_{1}\sigma_{4}}(x_4)&
\Psi_{\beta_3}^{\sigma_{4}\sigma_{3}}(x_3)
\Psi_{\beta_2}^{\sigma_{3}\sigma_{2}}(x_2)
\Psi_{\beta_1}^{\sigma_{2}\sigma_{1}}(x_1)
\Bigr\rangle = \nonumber \\
&=
\int_{\mathbb{S}}d\beta_{32}\;
C_{\beta_{4}|\beta_{32}\beta_{1}}^{\sigma_{4}\sigma_{2}\sigma_{1}}
C_{\beta_{32}|\beta_{3}\beta_{2}}^{\sigma_{4}\sigma_{3}\sigma_{2}}\;
\mathcal{F}^{t}(\Delta_{\beta_i},\Delta_{\beta_{32}},x_i) .
\end{aligned}\end{equation}
Using the fusion transformations
(\ref{transfost}), the equivalence of the factorisation in
the two channels can be rewritten:
\begin{equation}
\int_{\Sp}d\beta_{21}\;
C_{\beta_{4}|\beta_{3},\beta_{21}}^{\sigma_{4}\sigma_{3}\sigma_{1}}
C_{\beta_{21}|\beta_{2}\beta_{1}}^{\sigma_{3}\sigma_{2}\sigma_{1}}\;
F_{\beta_{21}\beta_{32}}\!\big[\,{}_{\be_4}^{\be_3}\;{}_{\be_1}^{\be_2}
\,\big]
 =%
C_{\beta_{4}|\beta_{32},\beta_{1}}^{\sigma_{4}\sigma_{2}\sigma_{1}}
C_{\beta_{32}|\beta_{3}\beta_{2}}^{\sigma_{4}\sigma_{3}\sigma_{2}}.
\label{pentagone boundary}
\end{equation}
By means of the pentagon equation (\ref{pentagone}) it easy to verify
that the following ansatz
\begin{equation}\label{ansatz}
C_{\beta_{3}|\beta_{2}\beta_{1}}^{\sigma_{3}\sigma_{2}\sigma_{1}}=%
\frac{g_{\beta_{3}}^{\sigma_{3}\sigma_{1}}}
     {g_{\beta_{2}}^{\sigma_{3}\sigma_{2}}
      g_{\beta_{1}}^{\sigma_{2}\sigma_{1}}}
F_{\sigma_{2}\beta_{3}}\!\big[
\,{}_{\si_3}^{\be_2}\;{}_{\si_1}^{\be_1}\,\big]
\end{equation}
yields a solution to (\ref{pentagone boundary}), as was noticed in
\cite{runkel}.
The coefficients $g_{\beta}^{\sigma_{2}\sigma_{1}}$ appearing are 
unrestricted by (\ref{pentagone boundary}). Additional information is needed
to determine them.

\subsection{Determination of the function $g$}\label{gcalc}

The boundary three point function 
$C_{\beta_{3}\beta_{2}\beta_{1}}^{\sigma_{3}\sigma_{2}\sigma_{1}}$ 
should be meromorphic w.r.t. the variables $\be_3,\be_2,\be_1$.
This far-reaching assumption can be motivated
in various ways: One may e.g. use arguments like 
those reviewed in section 3 of \cite{Te} concering the path integral 
for Liouville theory. These arguments
exhibit the analytic properties of correlation functions as a reflection
of the asymptotic behavior of the Liouville path integral measure
in the region $\phi\rightarrow -\infty$ where the interaction terms vanish. 
\footnote{An equivalent discussion can be carried out
in the framework of canonical quantization by considering the asymptotic
behavior of the Hamiltonian and its generalized eigenfunctions, 
see \cite[Section 11]{Te} for such a discussion in the case of 
Liouville theory without boundary, and \cite{tesch1} for the
basics of the corresponding treatment in the case of boundary conditions
like (\ref{bound}).}

Such considerations lead in particular to
the identification of the residues for the poles 
of $C_{\beta_{3}\beta_{2}\beta_{1}}^{\sigma_{3}\sigma_{2}\sigma_{1}}$
with certain correlation functions in free field theory, 
which generalize the so-called screening-charge constructions 
of \cite{dotsenko}.  The resulting prescription for the calculation of these 
residues was formulated in \cite{FZZ}. Most relevant for our purposes 
will be the observation that 
$C_{\beta_{3}\beta_{2}\beta_{1}}^{\sigma_{3}\sigma_{2}\sigma_{1}}$
has a pole with residue 1 if $\be_1+\be_2+\be_3=Q$: 
The relevant correlation functions in free field theory do not contain
any screening charges.
 
On the other hand, it seems worth observing that the fusion coefficients 
themselves are meromorphic functions of all six variables they depend on,
see \cite[Lemma 21]{PT2}.
This means that the function 
$C_{\beta_{3}\beta_{2}\beta_{1}}^{\sigma_{3}\sigma_{2}\sigma_{1}}$
that is given by the expression (\ref{ansatz}) will be meromorphic
iff the function $g_{\be}^{\si_2\si_1}$ is meromophic w.r.t. $\be$.

In the following, we shall consider the special boundary field
$\Psi_{-b}^{\sigma\sigma}(x)$,
which corresponds to a degenerate representation of the Virasoro 
algebra. As pointed out in \cite{FZZ}, it is in general not a trivial issue
to decide when a boundary field that corresponds to 
a degenerate representation will  satisfy the corresponding
differential equations expressing null vector decoupling.
Here, however, one may observe that one may create 
the boundary field $\Psi_{-b}^{\sigma\sigma}(x)$ by sending 
the bulk field $V_{-b/2}$ to the boundary: It follows from the 
fact that $V_{-b/2}$ satisfies a second order differential equation 
that the asymptotic behavior when $V_{-b/2}$ approaches the 
boundary is described by a boundary field 
$\Psi_{-b}^{\sigma\sigma}(x)$ that satisfies a third order 
differential equation. This last fact also implies that the 
operator product expansion of $\Psi_{-b}^{\sigma\sigma}(x)$
with a generic boundary operator can only contain three types
of contributions:
\begin{equation}\begin{aligned}
{}& \Psi_{\beta}^{\sigma_{2}\sigma_{1}}(x')
\Psi_{-b}^{\sigma_1\sigma_1}(x) \;=\; \\
& \quad =
\sum_{s=-1}^{1}\;c_s(\be;\si_2;\si_1)|x'-x|^{\De_{\be-sb}-\De_\be-\De_{-b}}  
\Psi_{\beta -sb}^{\si_2\sigma_{1}}(x)+\;( {\rm descendants}). 
\label{OPE -b}
\end{aligned}\end{equation}
One may then consider the vacuum expectation values
of the product of operators that is obtained
by multiplying (\ref{OPE -b}) with the boundary fields
$\Psi_{Q-\beta+sb}^{\sigma_{1}\sigma_{3}}$, $s\in\{-,0,+\}$.
Taking into account (\ref{twopt}), one is led to identify
the structure functions $c_{s}(\beta;\si_2;\si_1)$ ($s=+,0,-)$ with 
residues of the general three point function. As mentioned 
previously, the relevant residues can  
be represented as correlation functions in free field theory.
The structure function $c_+$ is nothing but a 
special case of the above-mentioned
residue at $\be_1+\be_2+\be_3=Q$, which is 1.

This should be compared to what would follow from our ansatz
(\ref{ansatz}). Let us note that it follows from appendix B(iii)
that the fusion coefficients indeed have a pole in the presently 
considered case. The corresponding residue is most easily calculated 
by means of the recursion relations that follow from 
(\ref{pentagone}), see appendix B(iii) for some details. We find
\begin{equation}\begin{aligned}
{} F_{\sigma_{1},\beta_{2}-b}\; 
\big[\,{}_{\si_3}^{\be_2}\;{}_{\phantom{\cdot}\si_1}^{-b}\,\big]
 = \frac{\Gamma(1+b^{2})}{\Gamma(1+2b^{2})} & 
\frac{\Gamma(2b\sigma_1)\Gamma(2b(Q-\sigma_1))}{\Gamma(b(Q-\beta_2+\sigma_3-\sigma_1))\Gamma(b(Q-\beta_2+\sigma_1-\sigma_3))}\ti \\
\ti & \frac{\Gamma(b(Q-2\beta_2))\Gamma(b(Q-2\beta_2 +b))}{\Gamma(b(\sigma_3+\sigma_1-\beta_2))\Gamma(b(2Q -\beta_2 -\sigma_3-\sigma_1))}. 
\end{aligned}\end{equation}
Our ansatz (\ref{ansatz}) together with $c_+\equiv 1$ therefore implies 
the following 
first order difference equation for $g_{\beta_{2}}^{\sigma_{3}\sigma_{1}}$:
\begin{equation}\label{findiffg}
1=\frac{g_{\beta_{2}-b}^{\sigma_{3}\sigma_{1}}}
       {g_{\beta_{2}}^{\sigma_{3}\sigma_{1}}
        g_{-b}^{\sigma_{1}\sigma_{1}}}
  F_{\sigma_{1},\beta_{2}-b}\; 
\big[\,{}_{\si_3}^{\be_2}\;{}_{\phantom{\cdot}\si_1}^{-b}\,\big].
\end{equation}
This functional equation is solved by the following expression:
\begin{equation}\label{gexp}\boxed{\;\;\begin{aligned}
{} g_{\beta}^{\sigma_3\sigma_1}\,=\, 
&\frac{\big(\pi \mu \gamma(b^{2})b^{2-2b^{2}}\big)^{\beta/2b}}
{\Gamma_{b}(2Q-\beta-\sigma_{1}-\sigma_{3})}\\
\ti& 
\frac{\Gamma_{b}(Q)\Gamma_{b}(Q-2\beta)\Gamma_{b}(2\sigma_{1})
\Gamma_{b}(2Q-2\sigma_{3})}{\Gamma_{b}(\sigma_{1}+\sigma_{3}-\beta)
\Gamma_{b}(Q-\beta+\sigma_{1}-\sigma_{3})
\Gamma_{b}(Q-\beta+\sigma_{3}-\sigma_{1})}.
\end{aligned}\;\;}\end{equation}
In order to discuss the uniqueness of our solution (\ref{gexp}) let us
note that one may derive a second finite difference equation 
that is related to (\ref{findiffg}) by substituting $b\rightarrow b^{-1}$
if one considers $\Psi_{-b^{-1}}^{\sigma_1\sigma_1}$ instead of
$\Psi_{-b}^{\sigma_{1}\sigma_{1}}$. Taken together, these two
functional equations allow one to conclude that our solution 
(\ref{gexp}) is unique up to multiplication by a factor of the
form $(f(\si_1,\si_3))^{\beta/2b}$, 
at least for irrational values of $b$. 

To fix the remaining freedom 
it is useful to note that we now have two possible ways 
to calculate the structure function $c_-(\be;\si_2,\si_1)$:
On the one hand one may use 
our ansatz (\ref{ansatz}) together with (\ref{gexp}) and the following
residue of the fusion coefficients:
\[\begin{aligned}
{} & F_{\sigma_{1},\beta_{2}+b}\; 
\big[\,{}_{\si_3}^{\be_2}\;{}_{\phantom{\cdot}\si_1}^{-b}\,\big]
=\frac{\Gamma(1+b^{2})}{\Gamma(1+2b^{2})}. \\
& \cdot\frac{\Gamma(2b\sigma_1)\Gamma(2b(Q-\sigma_1))\Gamma(2b\beta_2-2bQ)\Gamma(2b\beta_2-1)}{\Gamma(b(\beta_2+\sigma_3-\sigma_1))\Gamma(b(\beta_2+\sigma_1-\sigma_3))\Gamma(b(\sigma_3+\sigma_1+\beta_2 -Q))\Gamma(b(\beta_2 -\sigma_3-\sigma_1+Q))} 
\end{aligned}\]
On the other hand, $c_-(\be;\si_2,\si_1)$ 
is one of the cases where a representation in terms of free field 
correlation functions is available\cite{FZZ}:
\begin{equation}\begin{aligned}
c_-  (\be;\si_2,\si_1)\;=\;
  -& \frac{4\mu}{\pi}\frac{\Gamma(1+b^2)}{\Gamma(-b^2)}\\ 
& \ti\Gamma(b(2\beta_2-Q))\Gamma(2b\beta_2 -1)
\Gamma(1-2b\beta_2)\Gamma(1-b(2\beta_2+b))\\
& \ti\sin\pi b(Q+\beta_2-\sigma_3-\sigma_1) 
\sin\pi b(\beta_2 +\sigma_3+\sigma_1-Q) \\
& \ti\sin\pi b(\beta_2+\sigma_3-\sigma_1) 
\sin\pi b(\beta_2 +\sigma_1-\sigma_3).
\end{aligned}\end{equation}
One finds a precisce coincidence of 
the expressions which one obtains by following these 
two ways if and only if the prefactor in the expression 
for $g_{\beta}^{\sigma_3\sigma_1}$ is the one chosen in (\ref{gexp}).

By collecting the pieces, one finally arrives at the 
following expression for the three point function of boundary operators:
\begin{equation}
\boxed{\;\begin{aligned}
{}  C_{\beta_{3}|\beta_{2}\beta_{1}}^{\sigma_{3}\sigma_{2}\sigma_{1}} =&
\bigl(\pi \mu \gamma(b^2) b^{2-2b^2}\bigr)^{\frac{1}{2b}(\beta_3-\beta_2-\beta_1)}\Gamma_b(2Q-\beta_1-\beta_2-\beta_3)\\
& \ti\frac{\Gamma_b(\beta_2+\beta_3-\beta_1)
 \Gamma_b(Q+\beta_2-\beta_1-\beta_3)\Gamma_b(Q+\beta_3-\beta_1-\beta_2)}
{\Gamma_b(2\beta_3-Q)\Gamma_b(Q-2\beta_2)\Gamma_b(Q-2\beta_1)\Gamma_b(Q)} \\
& \ti \frac{S_b(\beta_3+\sigma_1-\sigma_3)S_b(Q+\beta_3-\sigma_3-\sigma_1)}{S_b(\beta_2+\sigma_2-\sigma_3)S_b(Q+\beta_2-\sigma_3-\sigma_2)}  
\int\limits_{-\infty}^{\infty}\! ds \;\prod_{k=1}^4
\frac{S_b(U_k+is)}{S_b(V_k+is)}, 
\label{formule fonction 3 points}
\end{aligned}\;}\end{equation}
where the coefficients $U_i$, $V_i$ and $i=1,\ldots,4$ are defined as
$$
\begin{array}{ll}
 U_1 =\sigma_1+\sigma_2-\beta_1 ,           &  V_1 = Q+\sigma_2-\sigma_3-\beta_1+\beta_3, \\
 U_2 = Q-\sigma_1+\sigma_{2}-\beta_1,       &  V_2 = 2Q+\sigma_2-\sigma_3-\beta_1-\beta_3, \\
 U_3 = \beta_2+\sigma_2-\sigma_3,           &  V_3 = 2\sigma_2, \\
 U_4 = Q-\beta_2+\sigma_2-\sigma_3.     & V_4=Q    
\end{array}
$$

\subsection{Further consistency checks} 

\begin{enumerate}
\item
One recovers the expression for the boundary reflection amplitude 
(\ref{reflcoeff}) from the
boundary three point function the same way the bulk reflection
amplitude was recovered from the bulk three point function in
\cite{AAl}: Using the fact that the fusion matrix depends on
conformal weights only, and is thus invariant when $\beta_{i}
\rightarrow Q-\beta_{i}$, one finds:
\begin{equation}
C_{Q-\beta_{3}|\beta_{2}\beta_{1}}^{\sigma_{3}\sigma_{2}\sigma_{1}} =%
\frac{g_{Q-\beta_{3}}^{\sigma_{3}\sigma_{1}}}
     {g_{\beta_{3}}^{\sigma_{3}\sigma_{1}}}
C_{\beta_{3}|\beta_{2}\beta_{1}}^{\sigma_{3}\sigma_{2}\sigma_{1}}
\end{equation}
From the expression (\ref{gexp}) for the function $g$, one 
indeed finds formula (\ref{reflcoeff}) for $S(\be;\si_2,\si_1)$.
\item
One may explicitly check that the two-point function (\ref{twopt})
is recovered by taking e.g. the limit $\beta_1 \to 0$ if the 
three-point function:
$$\lim_{\beta_1 \to 0}C_{\beta_{3}|\beta_{2}\beta_{1}}^{\sigma_{3}\sigma_{2}\sigma_{1}} 
\;=\; \delta (\beta_3 - \beta_2)+S(\be_2;\si_3\,\si_1)\de(\be_3+\be_2-Q).
$$
This is an easy consequence of the identity
(\ref{limfusion1}) proven in Appendix B(i)
\item
With the help of 
symmetry properties of the fusion coefficients
% \footnote{Details will be given elsewhere.} 
(see Appendix B(ii)), it is 
possible to check that the boundary three point function is 
invariant w.r.t. cyclic permutations.
\end{enumerate}

\subsection{Uniqueness}

We are finally going to sketch an argument in favor of the uniqueness of our 
expression for the boundary three point function:
Let us consider the associativity condition in the case that 
$\si_1=\si_2$ and that the boundary 
field $\Psi_{\be_1}^{\sigma_1\sigma_1}(x_1)$ is replaced by 
the degenerate field $\Psi_{-b}^{\sigma_1\sigma_1}(x_1)$.
Due to (\ref{OPE -b}), one finds that the associativity condition
(\ref{pentagone boundary}) gets replaced by
\begin{equation}
\sum_{\substack{\be_{21}=\be_2-sb\\ s\in\{-,0,+\}}}\;
c_s(\be_2;\si_3,\si_1)\,
F_{\beta_{21}\beta_{32}}\;
\big[\,{}_{\be_4}^{\be_3}\;{}_{-b}^{\phantom{.}\be_2}\,\big]
\,C_{\beta_{4}|\beta_{3}\beta_{21}}^{\sigma_{4}\sigma_{3}\sigma_{1}}
=\;
c_t(\be_{32};\si_4,\si_1)\;
C_{\beta_{32}|\beta_{3}\beta_{2}}^{\sigma_{4}\sigma_{3}\sigma_{1}},
\label{degasso}
\end{equation}
where $\be_{32}$ takes the values $\be_4+tb$, $t\in\{-,0,+\}$.
This can be read as a system of finite difference equations
for the general boundary three point function 
$C_{\beta_3|\beta_{2}\beta_{1}}^{\sigma_{3}\sigma_{2}\sigma_{1}}$. 
By specializing to the case that $\be_{32}=\be_4$, one finds in particular
a linear relation between the  
$C_{\beta_{4}|\beta_{3}\beta_{2}-sb}^{\sigma_{4}\sigma_{3}\sigma_{1}}$, 
$s\in\{-,0,+\}$. Replacing the degenerate field 
$\Psi_{-b}^{\sigma_1\sigma_1}(x_1)$ by 
$\Psi_{-b^{-1}}^{\sigma_1\sigma_1}(x_1)$ leads to  
a similar second order finite difference equation which is related to the
first by replacing $b\ra b^{-1}$ in the coefficients, as well as replacing 
$\beta_{2}-sb$ by $\beta_{2}-sb^{-1}$.

It can be 
shown (see appendix C) that such self-dual systems of finite difference 
equations can for irrational $b$ only 
have at most two linearly independent solutions. The relevant linear 
combination of these two solutions can be fixed e.g. by imposing the 
correct behavior w.r.t. the reflection $\be_2\ra Q-\be_2$
as given by (\ref{reflprop}).
In this way one arrives at the conclusion that the
finite difference equations following from the associativity condition
together with the reflection property (\ref{reflprop}) suffice to 
uniquely determine the dependence of 
$C_{\beta_{3}|\beta_{2}\beta_{1}}^{\sigma_{3}\sigma_{2}\sigma_{1}}$
w.r.t. the variable $\be_1$. 

But one may of course repeat that line of arguments by replacing any
of the four operators in the four point function of boundary fields
by the degenerate fields 
$\Psi_{-b}^{\sigma\sigma}(x)$ or 
$\Psi_{-b^{-1}}^{\sigma\sigma}(x)$, which would lead
to finite difference equations that constrain the dependence 
of  $C_{\beta_{3}|\beta_{2}\beta_{1}}^{\sigma_{3}\sigma_{2}\sigma_{1}}$
w.r.t. $\be_3$ and $\be_2$. This leads to the conclusion that 
indeed the associativity condition in the presence of degenerate fields 
together with the reflection property (\ref{reflprop}) uniquely 
determine the dependence of 
$C_{\beta_{3}|\beta_{2}\beta_{1}}^{\sigma_{3}\sigma_{2}\sigma_{1}}$
w.r.t. all three variables $\be_3,\be_2,\be_1$.

The remaining freedom consists of multiplication with an arbitrary
function of the boundary parameters $\si_3,\si_2,\si_1$. This 
freedom is eliminated by requiring that the residue of the
pole of $C_{\beta_{3}|\beta_{2}\beta_{1}}^{\sigma_{3}\sigma_{2}\sigma_{1}}$
at $\be_1+\be_2+\be_3=Q$ should indeed be unity, as discussed in subsection
\ref{gcalc}.

\section{Concluding remarks}

We now have determined the last of the structure functions that
one needs to 
completely characterize Liouville theory on domains with boundary.
What remains to be done is the verification that the expressions that
have been put forward indeed satisfy the full set of Cardy-Lewellen type
\cite{CL} consistency conditions. Although some particularly important
conditions have been verified (an analog of the Cardy condition \cite{tesch1},
as well as the associativity condition studied in the present paper),
it remains in particular to verify the conditions that link 
the boundary three point function with the bulk-boundary two-point function
proposed in \cite{hosomichi}.

A beautiful characterization of the structure constants of certain
classes of rational conformal field theories with boundaries has been
given in \cite{felder}, see also \cite{BPPZ} for closely related results. 
It can be read as the statement that upon 
choosing a suitable normalization of the three point conformal blocks or
chiral vertex operators, it becomes possible to recover all of the 
structure constants from the defining data of an associated modular 
tensor category. Validity of the Cardy-Lewellen conditions 
is automatic in this formalism. What is not directly
furnished by that formalism, though, is the 
explicit characterization for the necessary
normalization of the three point conformal blocks. 

It would certainly be nice to have at hand a similarly powerful formalism
for non-rational conformal field theories such as Liouville theory. 
This should allow one in particular to carry out the missing proof
that the structure functions satisfy the full set of Cardy-Lewellen type
\cite{CL} consistency conditions. We will therefore try to verify whether
our expression for the boundary three point function can be written
in a form that one would expect to find in an extension of the 
formalism of \cite{felder} to non-rational CFT. 

This turns out to be the case:
Let us write the three point function in terms of the
b-Racah-Wigner coefficients:
\begin{equation}
C_{\beta_{3}|\beta_{2}\beta_{1}}^{\sigma_{3}\sigma_{2}\sigma_{1}}\;=\;
\frac{g_{\beta_{3}}^{\sigma_{3}\sigma_{1}}}
{g_{\beta_{2}}^{\sigma_{3}\sigma_{2}}g_{\beta_{1}}^{\sigma_{2}\sigma_1}}
\frac{N(\sigma_3,\beta_2,\sigma_2) N(\sigma_2,\beta_1,\sigma_1)}{N(\sigma_3,\beta_3,\sigma_1) N(\beta_{3},\beta_2,\beta_1)}\my6j{\sigma_1}{\beta_2}{\beta_1}{\sigma_3}{\sigma_{2}}{\beta_{3}} 
\end{equation}
where \cite{PT1}
\begin{equation}\begin{aligned}
{}& N( \beta_3  ,\beta_2,\beta_1)
\;=\; \\
 &   \frac{\Ga(2\beta_1)\Ga(2\beta_2)\Ga(2Q-2\beta_3)}
{\Ga(2Q-\beta_1-\beta_2-\beta_3)\Ga(Q-\beta_1-\beta_2+\beta_3)
 \Ga(\beta_1+\beta_3-\beta_2)\Ga(\beta_2+\beta_3-\beta_1)}
\end{aligned}\end{equation}
It is easy to see that this can be rewritten as 
\begin{equation}\label{C-6j}
C_{\beta_{3}|\beta_{2}\beta_{1}}^{\sigma_{3}\sigma_{2}\sigma_{1}}=%
\big(g_{\beta_{3}}^{\beta_2\beta_1}\big)^{-1}
\new6j{\sigma_1}{\beta_2}{\beta_1}{\sigma_3}{\sigma_{2}}{\beta_{3}}, 
\end{equation}
where the b-Racah-Wigner coefficients that appear on the right hand side
have been modified w.r.t. to those considered in \cite{PT2} according to  
\begin{equation}\begin{aligned}
{}\new6j{\sigma_1}{\beta_2}{\beta_1}{\sigma_3}{\sigma_{2}}{\beta_{3}}\;\equiv\;
\frac{S_b(\si_3+\be_3-\si_1)S_b(\be_3+\be_2-\be_1)}
{S_b(\si_3+\be_2-\si_2)S_b(\si_2+\be_1-\si_1)}
\my6j{\sigma_1}{\beta_2}{\beta_1}{\sigma_3}{\sigma_{2}}{\beta_{3}}.
\end{aligned}\end{equation}
By using the the counterpart of (\ref{crossid}) for the b-Racah-Wigner
coefficients one may write (\ref{C-6j}) as
\begin{equation}\label{C-6j'}
C_{\beta_{3}|\beta_{2}\beta_{1}}^{\sigma_{3}\sigma_{2}\sigma_{1}}=%
\frac{1}{g(\beta_{3};\beta_2,\beta_1)}
\new6j{\bar{\beta}_2}{\sigma_1}{\bar{\beta}_1}{\sigma_3}
{\bar{\beta}_{3}}{\sigma_{2}}. 
\end{equation}
We consider (\ref{C-6j'}) as an encouraging hint that a verification
of the Cardy-Lewellen conditions should be possible along similar
lines as followed in \cite{felder} for the case of rational
CFT.

\section{Acknowledgments}
This work was done while B.P. was at Laboratoire de Physique 
Th\'eorique et Math\'ematique, Universite Montpellier II, 
Montpellier, France, as part of the TMR network with contract
ERBFMRX CT960012. B.P. now acknowledges support by the
GIF-Project under contract Nr I-645-130-14/1999. 
J.T. is supported by DFG SFB 288 ``Differentialgeometrie 
und Quantenphysik''.
\newpage

\appendix
\section{Special functions}

\subsection{The function $\Gamma_b(x)$}

The function $\Gamma_b(x)$ is a close relative of the double
Gamma function studied in \cite{Ba,Sh}. It 
can be defined by means of the integral representation
\begin{equation}
\log\Gamma_b(x)\;=\;\int\limits_0^{\infty}\frac{dt}{t}
\biggl(\frac{e^{-xt}-e^{-Qt/2}}{(1-e^{-bt})(1-e^{-t/b})}-
\frac{(Q-2x)^2}{8e^t}-\frac{Q-2x}{t}\biggl)\;\;.
\end{equation}
Important properties of $\Gamma_b(x)$ are
\begin{align}
{}& \text{(i) Functional equation:} \quad
\Gamma_b(x+b)=\sqrt{2\pi}b^{bx-\frac{1}{2}}\Gamma^{-1}(bx)\Gamma_b(x). \label{Ga_feq}\\
{}& \text{(ii) Analyticity:}\quad
\Gamma_b(x)\;\text{is meromorphic,}\nonumber\\ 
{}& \hspace{2.5cm}\text{poles:}\;\,  
x=-nb-mb^{-1}, n,m\in\BZ^{\geq 0}.\\
{}& \text{(iii) Self-duality:}\quad \Gamma_b(x)=\Gamma_{1/b}(x). \label{self-dual} 
\end{align}

\subsection{The function $S_b(x)$}

The function $S_b(x)$ may be defined in terms of 
$\Gamma_b(x)$ as follows
\begin{equation}\label{sbdef}
S_b(x)\;=\;\Gamma_b(x)\,/\,\Gamma_b\big(Q-x)\;.
\end{equation}
% This function, or close relatives of it like 
% $e_b(x)\;=\;e^{\frac{\pi i}{2}x^2}\,e^{-\frac{\pi i}{24}(2-Q^2)}S_b(x)$,
% have appeared in the literature under various names like 
%``Quantum Dilogarithm'' \cite{FK1}, ``Hyperbolic G-function''
% \cite{Ru}, ``Quantum Exponential Function'' \cite{W} 
% and ``Double Sine Function'', we refer to 
% the appendix of \cite{KLS} for
% a useful collection of properties of $S_b(x)$ and further references.
An integral that represents $\log S_b(x)$ is
\begin{equation}
\log S_b(x)\;=\;\int\limits_0^{\infty}\frac{dt}{t}
\biggl(\frac{\sinh t(Q-2x)}{2\sinh bt\sinh b^{-1}t}-
\frac{Q-2x}{2t}\biggl)\;\;.
\end{equation}
The most important properties for our purposes are 
\begin{align}
{}& \text{(i) Functional equation:} \quad
S_b(x+b)\;=\;2\sin \pi b x\;
S_b(x). \label{sb_feq}\\
{}& \text{(ii) Analyticity:}\quad
S_b(x)\;\text{is meromorphic,}\nonumber\\ 
{}& \hspace{2.7cm}\text{poles:}\;\,  
x=-(nb+mb^{-1}), n,m\in\BZ^{\geq 0}.\\
{}& \hspace{2.7cm}\text{zeros:}\;\,  
x=Q+(nb+mb^{-1}), n,m\in\BZ^{\geq 0}.\nonumber \\
{}& \text{(iii) Self-duality:}\quad S_b(x)=S_{1/b}(x).  \\
{}& \text{(iv) Inversion relation:}\quad S_b(x)S_b(Q-x)\;=\;1.\\
{}& \text{(v) Asymptotics:} \quad S_b(x)\sim e^{\mp\frac{\pi i}{2}x(x-Q)}\;\,{\rm for}\;{\rm Im}(x)\ra\pm\infty\\
{}& \text{(vi) Residue:} \quad {\rm res}_{x=c_b}S_b(x)
=(2\pi)^{-1}\label{sbRes}.
\end{align}

\subsection{$\up$ function}
The $\up$ may be defined in terms of
$\Gamma_b$ as follows 
\begin{equation}
\up(x)^{-1} \equiv \Gamma_b(x)\Gamma_b(Q-x)\;.
\end{equation}
An integral representation convergent in the strip $0<{\rm Re}(x)<Q$ is 
\begin{eqnarray}
&&\text{log}\up(x)=\int_{0}^{\infty}\frac{dt}{t}\left\lbrack\left(\frac{Q}{2}-x\right)^{2}e^{-t}-\frac{\text{sinh}^{2}(\frac{Q}{2}-x)\frac{t}{2}}{\text{sinh}\frac{bt}{2}\text{sinh}\frac{t}{2b}}\right\rbrack\;. \nonumber 
\end{eqnarray}
Properties:
\begin{align}
{}& \text{(i) Functional equation:} \quad
\up(x+b)\;=\;\frac{\Ga(bx)}{\Ga(1-bx)}b^{1-2bx}\;
\up(x). \label{up_feq}\\
{}& \text{(ii) Analyticity:}\quad
\up(x)\;\text{is entire analytic,}\nonumber\\ 
{}& \hspace{2.7cm}\text{zeros:}\;\,  
x=-(nb+mb^{-1}), n,m\in\BZ^{\geq 0}.\\
{}& \hspace{2.7cm}\phantom{\text{zeros:}}\;\,  
x=Q+(nb+mb^{-1}), n,m\in\BZ^{\geq 0}.\nonumber \\
{}& \text{(iii) Self-duality:}\quad \up(x)=\Upsilon_{1/b}(x).
\end{align}

\section{Useful properties of the fusion coefficients}

\subsection{Some limiting cases of the fusion coefficients}

In this appendix we shall consider two important limiting cases of the 
fusion coefficients. We are going to show:
\begin{itemize}
\item[i)] If 
$\al_3=\frac{Q}{2}+iP_3$, $\al_t=\frac{Q}{2}+iP_t$ then
\begin{equation}\label{limfusion1}
\lim_{\al_2\ra 0}\;
\fus{\al_1}{\al_2}{\al_3}{\al_4}{\al_{1}}{\al_{t}}\;=\;
\de(P_t-P_3). 
\end{equation}
\item[ii)] Introduce $\tilde{C}(\al_3,\al_2,\al_1)\equiv
(\pi \mu \gamma(b^{2})b^{2-2b^{2}})^{\frac{1}{b}(\sum_{i=1}^3\al_i-Q)}
\Upsilon_0^{-1}
C(\al_3,\al_2,\al_1)$. Then:
\begin{equation}
\lim_{\al_s\ra 0}\fus{\al_1}{\al_1}{\al_2}{\al_2}{\al_{s}}{\al_{t}}\;=\;
\frac{\Ga(2Q)}{\Ga(Q)}\frac{S_b(2\al_t)}{S_b(2\al_t-Q)}
\tilde{C}(\al_3,\al_2,\al_1).
\end{equation}
\end{itemize}

To prove i), we will study the distribution on 
$\mathcal{S}'(\mathbb{R}\times\mathbb{R})$
defined as
\begin{equation}
\begin{aligned}
I_{\sigma_3,\sigma_1} & (p_3,p_2)\equiv  \\
& \equiv\lim_{\beta_1\to 0}\frac{1}{i}
\int\limits_{-i\infty}^{i\infty}ds \;
\frac{S_b(U_1+s)S_b(U_2+s)S_b(U_3+s)S_b(U_4+s)}
{S_b(V_1+s)S_b(V_2+s)S_b(V_3+s)S_b(Q+s)}
\Biggr|^{\sigma_1=\sigma_2}_{\beta_j=\frac{Q}{2}+ip_j;\;j=2,3.}
\nonumber
\end{aligned}
\end{equation}
It should be remarked that in sending $\beta_1\to 0$
some of the poles at $V_1+s=Q+nb$ and $V_2+s=Q+nb$
will cross the imaginary axis so that one has to
deform the contour accordingly.
If one first considers $I_{\sigma_3,\sigma_1}(p_3,p_1)$ for $p_1\neq p_3$
one finds by changing the integration variable to
$t=\sigma_1-\sigma_3+\beta_3+s$ that the integral simplifies to
a special value of the $b$-hypergeometric function:
\begin{equation}
\begin{aligned}
\int\limits_{-i\infty}^{i\infty}dt \; &
\frac{S_b(\beta_2-\beta_3+t)S_b(Q-\beta_2-\beta_3+t)}{S_b(2Q-2\beta_3+t)S_b(Q+t)}
=\\
& =\frac{S_b(\beta_2-\beta_3)S_b(Q-\beta_2-\beta_3)}{S_b(2Q-2\beta_3)}
F_b(\beta_2-\beta_3,Q-\beta_2-\beta_3;2Q-2\beta_3;0).
\nonumber
\end{aligned}
\end{equation}
This particular value of the $b$-hypergeometric function
vanishes, as follows from the identity \cite{PT2}
\begin{equation}
F_b(\al,\beta;\gamma;\frac{1}{2}(\gamma-\beta-\al-Q))=
e^{-2\pi i \al\beta}\frac{G_b(\gamma)G_b(\gamma-\al-\beta)}{G_b(\gamma-\al)G_b(\gamma-\beta)}\nonumber
\end{equation}
and the fact that $G_b(\gamma-\al-\beta)$ has a zero for $\gamma-\al-\beta=Q$.
One has thereby found that $I_{\sigma_3,\sigma_1}(p_3,p_2)$ has support
only for $\beta_3=\beta_2$.
In order to analyze the singular behavior
near $\beta_3=\beta_2$ it will be useful to split off the residue
contributions of the first poles that have crossed the real axis:
\begin{equation}
\begin{aligned}
 I_{\sigma_3,\sigma_1}  & (p_3,p_2)= I'_{\sigma_3,\sigma_1} (p_3,p_2)\\
-& \lim_{\epsilon \to 0}\biggl(
\frac{S_b(\beta_3+\beta_2-Q)S_b(\beta_3-\beta_2-\epsilon)}{S_b(2\beta_3)}
+\frac{S_b(\beta_2-\beta_3-\epsilon)S_b(Q-\beta_2-\beta_3)}{S_b(2Q-2\beta_3)}\biggr),
\nonumber
\end{aligned}
\end{equation}
where $I'_{\sigma_3,\sigma_1} (p_3,p_2)$ is defined by a contour that
passes to the right of the poles at $V_1+s=Q$ and $V_2+s=Q$.
One observes that $I'_{\sigma_3,\sigma_1} (p_3,p_2)$ is nonsingular at $\beta_3=\beta_2$,
and that $S_b(x)\sim \frac{1}{2\pi x}$ near $x=0$.
The singular behavior near $\beta_3=\beta_2$ is therefore given by
\begin{eqnarray}
-\frac{1}{2\pi}
\lim_{\epsilon \to 0}\Bigl( \frac{1}{i(p_3-p_2)-\epsilon}+\frac{1}{i(p_2-p_3)-\epsilon}
\Bigr)= 
\frac{1}{2\pi}\lim_{\epsilon \to 0}\frac{2\epsilon}{(p_3-p_2)^2+\epsilon^2}=\delta(p_3-p_2) \nonumber 
\end{eqnarray}
We have therefore found that
$I_{\sigma_3,\sigma_2} (p_3,p_2)=|S_b(2\beta_3)|^{-2}\delta(p_3-p_2)$.
Our claim i) is an easy consequence of this fact.

In order to verify ii), one should observe that the prefactor of the 
integral in the expression for the fusion coefficients vanishes.
However, the contour of integration gets pinched between
the poles from the factors $S_b(s+\al_s)$ and $S_b^{-1}(s+Q)$ of the 
integrand in taking the limit. To isolate the singular contribution 
of the integral, one may deform the contour $i\BR$ into a contour that
goes around the pole at $s=0$ in the right half plane plus a small circle
around $s=0$. Due to the vanishing prefactor, only the residue contribution
survives in the limit. The rest is straightforward.

\subsection{Symmetries of the fusion coefficients}

The fusion cofficients satisfy two types of symmetry
relations: First, one may permute pairs of the variables
$\al_1,\dots,\al_4$:
\begin{equation}
\fus{\al_1}{\al_2}{\al_3}{\al_4}{\al_{s}}{\al_{t}}= \fus{\al_4}{\al_3}{\al_2}{\al_1}{\al_{s}}{\al_{t}}= 
\fus{\al_3}{\al_4}{\al_1}{\al_2}{\al_{s}}{\al_{t}}. 
\end{equation}
These identities follow from similar identities for the
b-Racah Wigner symbols:
\begin{equation}\begin{aligned}
\my6j{\al_1}{\al_3}{\al_2}{\al_4}{\al_s}{\al_t}=&
\my6j{\al_3}{\al_1}{Q-\al_4}{Q-\al_2}{Q-\al_s}{Q-\al_t}\\
= & \my6j{\phantom{Q}\al_2}{Q-\al_4}{\phantom{Q}\al_1}{Q-\al_3}
         {\phantom{Q}\al_s}{Q-\al_t},
\end{aligned}\end{equation}
which are easily derived from the definition of the b-Racah Wigner symbols
given in \cite{PT2} taking into account the following properties of the
b-Clebsch-Gordan coefficients:
\newcommand{\CGC}[6]{\big[ \,{}^{#1}_{#2}\;{}^{#3}_{#4}\;{}^{#5}_{#6}\,
\big]_b}
\begin{equation}\begin{aligned}
\left(\CGC{\al_3}{x_3}{\al_2}{x_2}{\al_1}{x_1}\right)^*
=& e^{-\pi i\al_2^*(Q-\al_2^*)}
\big[\begin{smallmatrix} 
Q-\al_1^* & \al_2^* & Q-\al_3^*\\
x_1-c_b& {x_2} & x_3-c_b
\end{smallmatrix}\big]\\
=&
e^{+\pi i(\al_3^*(Q-\al_3^*)-\al_2^*(Q-\al_2^*)-\al_1^*(Q-\al_1^*))}
\CGC{\al_3^*}{x_3}{\al_1^*}{x_1}{\al_2^*}{x_2},
\end{aligned}\end{equation}
where we have used the notation $c_b=i\frac{Q}{2}$.
Second, there are identities that exchange the two ``internal indices''
with a pair of ``external indices''  
\begin{equation}\begin{aligned}
\fus{\al_1}{\al_2}{\al_3}{\al_4}{\al_{s}}{\al_{t}}
\fus{\al_{s}}{\al_{s}}{\al_3}{\al_3}{0}{\al_{4}}\;=\;& 
\fus{\al_{1}}{\al_s}{\al_{3}}{\al_t}{\al_{2}}{\al_{4}}
\fus{\al_2}{\al_2}{\al_3}{\al_3}{0}{\al_{t}}, \\
\fus{\al_1}{\al_2}{\al_3}{\al_4}{\al_{s}}{\al_{t}}
\fus{\al_1}{\al_{t}}{\al_{t}}{\al_1}{\al_{4}}{0}\;=\;& 
\fus{\al_{1}}{\al_t}{\al_{3}}{\al_s}{\al_{4}}{\al_{2}}
\fus{\al_1}{\al_2}{\al_2}{\al_1}{\al_{s}}{0}.\end{aligned}\end{equation}
The first of these identities is obtained from the pentagon 
(\ref{pentagone}) by setting $\be_1=\al_3$ and taking the limit 
$\be_2\ra 0$ with the help of (\ref{limfusion1}). The second can 
be obtained from the first by taking into account 
\begin{equation}\label{crossid}\begin{aligned}
C(\al_4,\al_3,\al_s)C(Q-\al_s,\al_2,\al_1)& 
\fus{\al_1}{\al_2}{\al_3}{\al_4}{\al_{s}}{\al_{t}}
=\\
=& C(\al_4,\al_t,\al_1)C(Q-\al_t,\al_3,\al_2) 
\fus{\al_3}{\al_2}{\al_1}{\al_4}{\al_{t}}{\al_{s}}.
\end{aligned}\end{equation}
This identity in turn follows via 
standard Moore-Seiberg type arguments 
\cite{moore} from the fundamental identity 
that assures crossing symmetry \cite{PT1}\cite{PT2}, together
with ($\al=\frac{Q}{2}+iP$,  $\al'=\frac{Q}{2}+iP'$)
\begin{equation}
\int_{\BS}d\be \;\, \fus{\al_1}{\al_2}{\al_3}{\al_4}{\al}{\be}
\fus{\al_3}{\al_2}{\al_1}{\al_4}{\be}{\al'}\; =\;\de(P-P').
\end{equation}

\subsection{Some residues of the fusion coefficients}

If one considers the special cases where one of $\al_1,\dots,\al_4$, say 
$\al_i$ equals $-\frac{n}{2}b-\frac{m}{2}b^{-1}$ and where a triple
$(\De_{\al_4},\De_{\al_3},\De_{\al_{21}})$, 
$(\De_{\al_{21}},\De_{\al_2},\De_{\al_1})$ which contains $\De_{\al_i}$
satisfies the fusion rules of \cite{FF}, one will find that
the right hand side of the fusion relation (\ref{transfost})
reduces to a finite sum of terms selected by the fusion rules of
\cite{FF}. The fusion coefficients that multiply the conformal 
blocks are residues of the general fusion coefficients, 
as can be seen by a generalization of our calculation
leading to (\ref{limfusion1}). 
In order to derive explicit expressions for these residues, it 
is useful to observe that the pentagon equation (\ref{pentagone}) 
leads to recursion relations that determines the above-mentioned 
residues in terms of the following special case:
\begin{eqnarray}
F_{s,s'}(\be|\si_1,\si_2) \;\equiv\;
F_{\sigma_{1}-\frac{sb}{2},\beta-\frac{s'b}{2}}\left[
\begin{smallmatrix}
\beta     & -\frac{b}{2}\\
\sigma_{2}    &   \phantom{r}\sigma_{1}%
\end{smallmatrix}
\right],
\nonumber
\end{eqnarray}
where $s,s' =\pm$. The explicit expressions for these coefficients are:
\begin{eqnarray}
F_{++}&=&\frac{\Gamma(b(2\sigma_1-b))\Gamma(b(b-2\beta)+1)}{\Gamma(b(\sigma_1-\beta-\sigma_2+b/2)+1)\Gamma(b(\sigma_1-\beta+\sigma_2-b/2))} \nonumber \\
F_{+-}&=& \frac{\Gamma(b(2\sigma_1-b))\Gamma(b(2\beta-b)-1)}{\Gamma(b(\sigma_1+\beta+\sigma_2-3b/2)-1)\Gamma(b(\sigma_1+\beta-\sigma_2-b/2))} \nonumber \\
F_{-+}&=&\frac{\Gamma(2-b(2\sigma_1-b))\Gamma(b(b-2\beta)+1)}{\Gamma(2-b(\sigma_1+\beta+\sigma_2-3b/2))\Gamma(1-b(\sigma_1+\beta-\sigma_2-b/2))} \nonumber \\
F_{--}&=& \frac{\Gamma(2-b(2\sigma_1-b))\Gamma(b(2\beta-b)-1)}{\Gamma(b(-\sigma_1+\beta+\sigma_2-b/2))\Gamma(b(-\sigma_1+\beta-\sigma_2+b/2)+1)} \nonumber 
\label{f1/2}
\end{eqnarray}
In subsection \ref{gcalc} we need the following fusion coefficients:
\begin{equation}
{} F_{\sigma_{1},\beta_{2}\pm b} \left[
\begin{smallmatrix}
\beta_{2}     & -b\\
\sigma_{3}    &   \phantom{r}\sigma_{1}%
\end{smallmatrix}
\right]
 = {} F_{\sigma_{1},\beta_{2}\pm b} \left[
\begin{smallmatrix}
  - b & \beta_{2} \\
\phantom{r}\sigma_{1} & \sigma_{3}      
\end{smallmatrix}
\right].
\end{equation}
The pentagon identity (\ref{pentagone}) then yields the formula that 
was used to calculate the expressions used in subsection \ref{gcalc}:
\begin{equation}
%\begin{aligned}
F_{\si_{2},\beta+sb} \left[
\begin{smallmatrix}
\beta     & -b\\
\sigma_{2}    &   \sigma_{1}%
\end{smallmatrix}
\right]\;=\;
\sum_{t=\pm}\;\frac{F_{t+}\big(-\fr{b}{2}|\be,\be+s\big)}
{F_{++}\big(-\frac{b}{2}|\si_2,\si_2\big)}\;
F_{-t}\big(\be|\si_2-\fr{b}{2},\si_1\big)
F_{+,s-t}\big(\be-\fr{tb}{2}|\si_2,\si_1\big)
.
%\end{aligned}
\end{equation}

\section{Uniqueness of solutions
of finite difference equations of the second order}

Let us indicate how one may obtain statements on the
uniqueness of such equations:
We will consider functions $f(x)$ that are analytic in some
domain $D$ that includes $i[0,2b^{-1}]$ and satisfy
\begin{equation}\label{diffeq}
\bigl(A_b(x)T^{2b} +\Psi_b(x)T^b +C_b(x)\bigr)f(x)=0,
\end{equation}
where $T^b$ is the operator defined by $T^bf(x)=f(x+b)$, 
as well as the difference equation obtained by replacing 
$b\ra b^{-1}$. We would like to 
show that there exist at most two linearly independent solutions.
Assume having three solutions $f_1$, $f_2$, $g$ of which $f_1$ and $f_2$
are linearly independent. One may consider  
\begin{equation}
\det\left|\begin{array}{ccc} 
g & f_1 & f_2 \\
T^bg & T^bf_1 & T^bf_2 \\
T^{2b}g & T^{2b}f_1 & T^{2b}f_2
\end{array}\right|.
\end{equation}
The determinant vanishes due to the fact that each row is a linear combination
of the two other by means of the difference equation (\ref{diffeq}). But this 
implies that also the columns must be linearly dependent, in particular
\begin{equation}
g\;=\;c_1(x)f_1+c_2(x)f_2,
\end{equation}
with coefficients $c_1$, $c_2$ that might a priori depend on $x$. These 
coefficients are found as 
\begin{equation}
c_1\;=\;\frac{\CW(g,f_2)}{\CW(f_1,f_2)},\qquad 
c_2\;=\;\frac{\CW(g,f_1)}{\CW(f_2,f_1)},\qquad
\CW(f,g)\;\equiv fT^bg-gT^bf.
\end{equation}
$\CW(f,g)$ can be seen as a q-analogue of the Wronskian relevant for
second order {it differential} equations. By a direct calculation using
(\ref{diffeq}) one finds that
\begin{equation}
T^b \;\CW(f,g)\;=\; \frac{C_b(x)}{A_b(x)}\CW(f,g) \quad \text{and}\quad
T^b \;c_i\;=\;c_i, \quad i=1,2. \end{equation} 
In a similar way 
one obtains $T^{1/b} c_i=c_i$, $i=1,2$. It then follows for irrational 
$b$ that the 
$c_i$, $i=1,2$ must be constant: From $T^b c_i=c_i$ one finds periodicity
of $d_i(x)\equiv c_i(-ix)$ in the interval $[0,2b]$, so that 
$c_i$ can be represented by a Fourier-series. One may then use the equation
$T^{1/b} c_i=c_i$ to show vanishing of all Fourier-coefficients but
the one of the constant mode.

\end{document}